\documentclass[twocolumn,prl]{revtex4}

\usepackage{graphicx}
\usepackage{dcolumn}
\usepackage{bm}
\usepackage{multirow}
\usepackage{color}
\usepackage[normalem]{ulem}
\usepackage{amsmath}
\usepackage{gensymb}
\usepackage[colorlinks=true]{hyperref}

\begin{document}

\title{Interlayer exciton mediated second harmonic generation in bilayer MoS$_2$}

\author{Shivangi Shree$^{1}$}
\author{Delphine Lagarde$^1$}
\author{Laurent Lombez$^1$}
\author{Cedric Robert$^1$}
\author{Andrea Balocchi$^1$}
\author{Kenji Watanabe$^2$}
\author{Takashi Taniguchi$^3$}
\author{Xavier Marie$^1$}
\author{Iann C. Gerber$^1$}
\author{Mikhail M. Glazov$^4$}
\email{glazov@coherent.ioffe.ru}
\author{Leonid E. Golub$^4$}
\author{Bernhard Urbaszek$^1$}
\email{urbaszek@insa-toulouse.fr}
\author{Ioannis Paradisanos$^{1}$}

\affiliation{$^1$Universit\'e de Toulouse, INSA-CNRS-UPS, LPCNO, 135 Avenue Rangueil, 31077 Toulouse, France}
\affiliation{$^2$Research Center for Functional Materials, National Institute for Materials Science, 1-1 Namiki, Tsukuba 305-0044, Japan}
\affiliation{$^3$International Center for Materials Nanoarchitectonics, National Institute for Materials Science,  1-1 Namiki, Tsukuba 305-0044, Japan}
\affiliation{$^4$Ioffe Institute, 194021 St.\,Petersburg, Russia}

\begin{abstract} Second harmonic generation (SHG) is a non-linear optical process, where two photons coherently combine into one photon of twice their energy. Efficient SHG occurs for crystals with broken inversion symmetry, such as transition metal dichalcogenide monolayers. Here we show tuning of non-linear optical processes in an inversion symmetric crystal. This tunability is based on the unique properties of bilayer MoS$_2$, that shows strong optical oscillator strength for the intra- but also interlayer exciton resonances. As we tune the SHG signal onto these resonances by varying the laser energy, the SHG amplitude is enhanced by several orders of magnitude. In the resonant case the bilayer SHG signal reaches amplitudes comparable to the off-resonant signal from a monolayer. In applied electric fields the interlayer exciton energies can be tuned due to their in-built electric dipole via the Stark effect. As a result the interlayer exciton degeneracy is lifted and the bilayer SHG response is further enhanced by an additional two orders of magnitude, well reproduced by our model calculations.
\end{abstract}


\maketitle

\textbf{Introduction.---} 
Non-linear optical spectroscopy of atomically thin crystals gives crucial insights into their physical properties and investigates potential applications \cite{zhang2020second,trovatello2020optical}. Non-linear optics is based on the interaction of photons within the crystal. This can give rise, among other processes, to second harmonic generation (SHG) where two photons of identical energy combine into one photon of twice this energy \cite{PhysRevLett.7.118,PhysRevLett.8.404}. Currently existing applications in photonics and laser physics make use of this process in bulk crystals for example using potassium dihydrogen phosphate (KDP) or lithium niobate (LiNbO$_3$)~\cite{yariv2007photonics}. \\
\indent SHG is vital both as a spectroscopic tool \cite{heinz1982spectroscopy,shree2020guide, fiebig2005second} and also for applications of a large class of materials such as 
II-VI and III-V semiconductors~\cite{lafrentz2013second,bergfeld2003second,Yakovlev:2018aa}, 
nanotubes \cite{doi:10.1063/1.1782255}, 
magnetic- and non-magnetic layered materials \cite{sun2019giant,lochner2019controlling,cunha2020second,seyler2015electrical, wang2015giant,lin2019quantum,mennel2018optical}, 
perovskites and antiferromagnetic oxydes  \cite{abdelwahab2018highly,doi:10.1063/1.367579}. \\
\indent  The occurrence of SHG in a crystal is directly linked to its symmetry, where crystals with broken inversion symmetry can exhibit SHG and crystals which are inversion symmetric can typically not \cite{yariv2007photonics}. This dependence on symmetry makes non-linear optics in atomically thin crystals such as graphene a very rich field of research \cite{Glazov2014101,doi:10.1063/1.3275740,hendry2010coherent,PhysRevB.84.045432,glazov2011second,gullans2013single,soavi2019hot}. Although pristine graphene is an inversion symmetric crystal, the crystal symmetry or symmetry of the electronic states is potentially very sensitive to charges or single molecules in the layer vicinity \cite{PhysRevLett.117.123904}, an imbalance in population of valley states \cite{golub2014valley}, or electric currents~\cite{PhysRevB.85.121413}. However, graphene is a gapless material, so light-matter interaction cannot be controlled to the same extent as in atom-thin semiconductors based on transition metal dichalcogenides (TMD). \\
\indent TMD monolayers such as MoS$_2$ and WSe$_2$ have broken inversion symmetry and show intrinsically strong SHG signals \cite{li2013probing,zhang2020second,PhysRevB.87.201401}. Excitons are strongly bound in these materials \cite{wang2018colloquium} and govern optical processes also at room temperature, contrary to
the model semiconductor GaAs \cite{bergfeld2003second}. This is crucial for non-linear optics as exciton resonances enhance light-matter interaction by several orders of magnitude \cite{wang2018colloquium}.  As a result, in monolayers the excitonic contribution to SHG can be orders of magnitude higher than the intrinsic contribution from the crystal which exists also off-resonance \cite{wang2015giant,seyler2015electrical,wang2015exciton,glazov2017intrinsic,zhao2016atomically}. An intrinsic challenge for exploiting SHG signals from nanostructures is the overall small signal, despite the giant response of a monolayer per unit thickness \cite{PhysRevB.87.201401,wen2019nonlinear,weismann2016theoretical}. Hence, in the interest of both fundamental physics and applications we need to uncover the exact origins of SHG in layered semiconducting materials to enable further tuning and amplification. \\
\begin{figure*}
\includegraphics[width=0.99\linewidth]{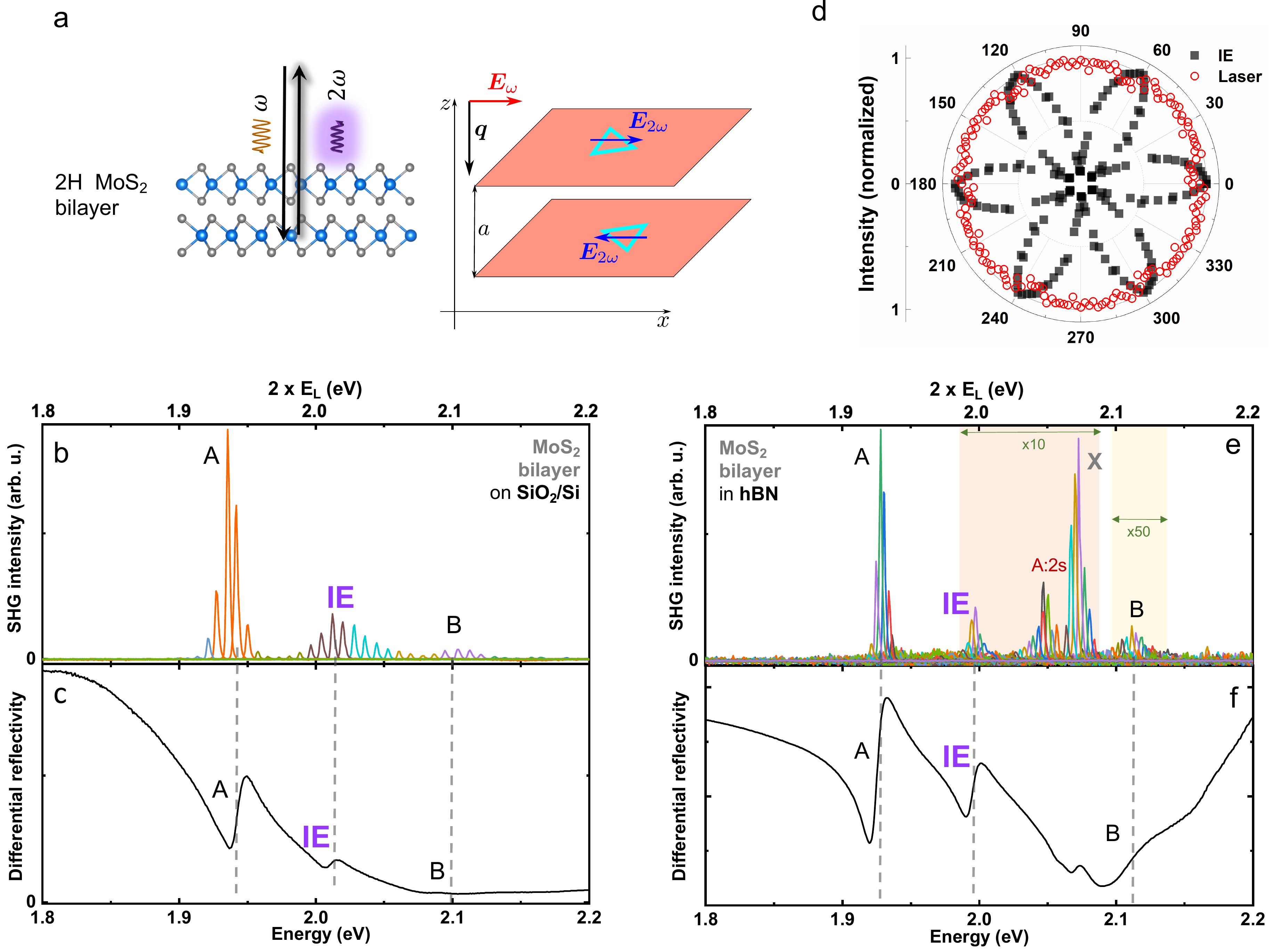}
\caption{\label{fig:fig1} \textbf{Second harmonic generation (SHG) spectroscopy in bilayer MoS$_2$ with 2H stacking in different dielectric environments.} \textbf{(a)}~Schematics of SHG in MoS$_2$ bilayers with 2H stacking. \textbf{(b)} SHG spectra plotted as a function of twice the laser energy $2E_L$. Each spectrum is a separate measurement and we plot all the SHG spectra for the different laser energies as in \cite{wang2015exciton}. SHG experiments are performed on non-encapsulated 2H MoS$_2$ bilayers (laser power 4 ~mW ) and  hBN-encapsulated samples (10~mW), see panel (e). The main excitonic transitions are marked: intralayer A, B and interlayer IE. \textbf{(c)} Differential reflectivity spectrum of the same sample as a function of photon energy. \textbf{(d)} Polarization-resolved SHG intensity \citep{li2013probing} for $2 \times E_L=$~ 2.01 eV in resonance with the interlayer exciton IE (black squares) energy of 2H MoS$_2$ bilayers. For comparison we show the laser reflection (red circles) which remains constant for each angle. \textbf{(e,f)} same as \textbf{(b,c)}, but for bilayer samples encapsulated in hBN. Note that in panel (e) we have multiplied the SHG signal in the orange shaded region by 10, and in the yellow shaded region by 50 to discern all transitions on a linear scale.}
\end{figure*}
\indent In this work we demonstrate strong, tunable SHG based on exciton resonances in nominally inversion symmetric MoS$_2$ bilayers with 2H stacking : Varying the laser excitation energy allows us to address resonantly not only intralayer and but also interlayer exciton states with large oscillator strength \citep{Leisgang2020}, which makes MoS$_2$ bilayers particularly interesting. At these specific energies we generate SHG signals in MoS$_2$ bilayers which are (i) comparable in amplitude to off-resonance SHG monolayer signals and (ii) several orders of magnitude larger than  reported in the literature for TMD bilayers with 2H stacking \cite{li2013probing,zhao2016atomically,Klein2017,yu2015charge}. 
The prominent role of interlayer excitons in MoS$_2$ bilayers allows for an additional SHG tuning mechanism: in applied electric fields we lift the degeneracy of the interlayer exciton states through the Stark effect. This results in an additional enhancement of the SHG signal at the interlayer exciton resonance by two orders of magnitude, surpassing the response of an ungated monolayer at this energy. We analyze the origin of the highly tunable SHG response through the linear polarization dependence of the SHG signal with respect to the crystallographic axis. We
compare bilayer results with measurements in monolayers and trilayers located in the same van der Waals stacks. We present a model analysis of the emergence of the strong SHG in the bilayer, taking into account the impact of electric fields perpendicular to the bilayer as well as different sources of asymmetry.\\
\begin{figure*}
\includegraphics[width=0.88\linewidth]{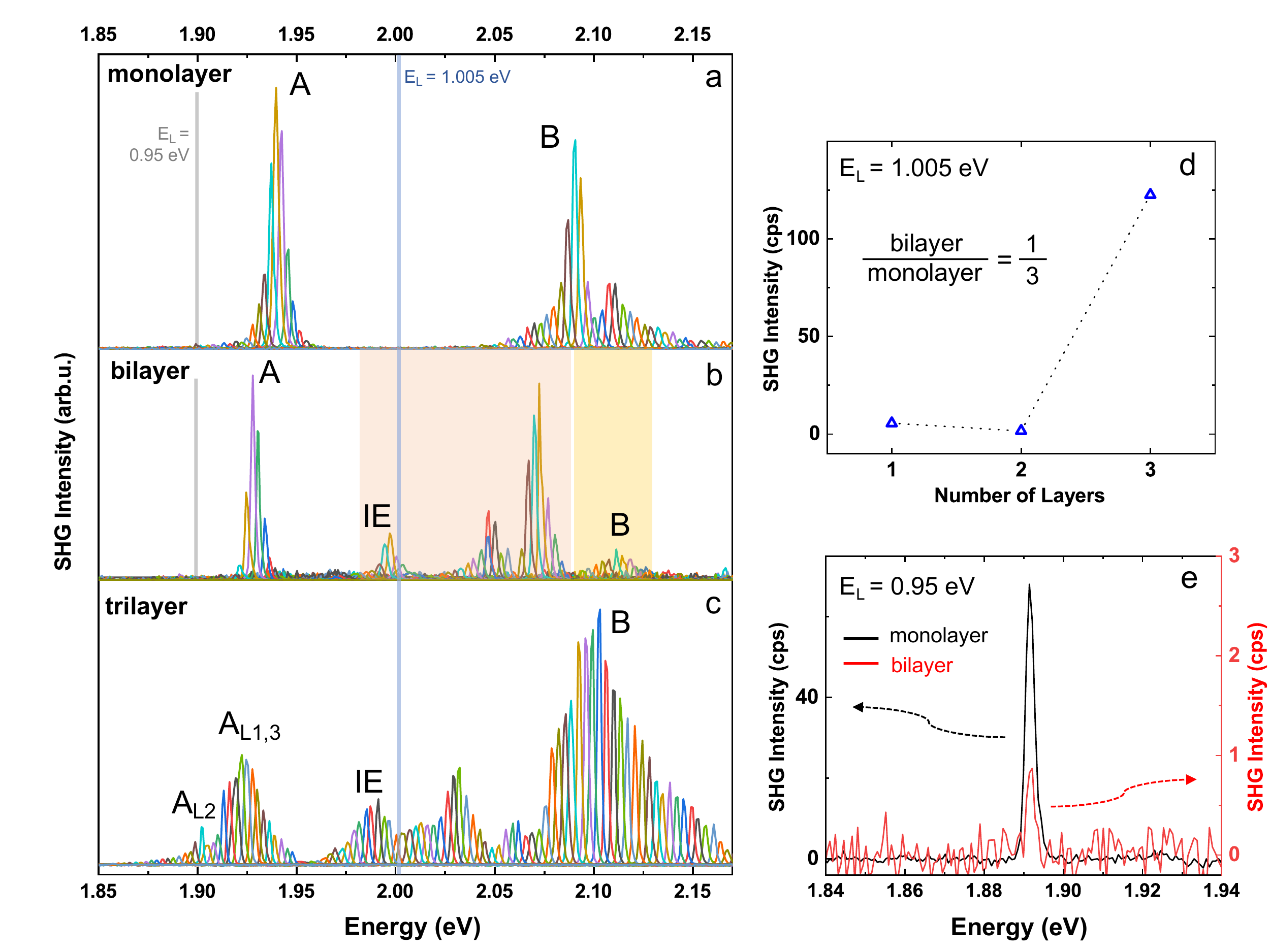}
\caption{\label{fig:fig2} \textbf{Comparing SHG spectroscopy for MoS$_2$ mono-, bi- and trilayers.} hBN encapsulated sample with staircase arrangement of MoS$_2$ mono-, bi- and trilayer. \textbf{(a)} SHG spectra plotted as a function of twice the laser excitation energy $2E_L$ for the monolayer. The main exciton transitions are marked, compare with Fig.~\ref{fig:fig1}e,f for the bilayer and the supplement for mono-and trilayers. \textbf{(b)}  SHG spectra for bilayer, where intensity in orange shaded region is $\times$10, yellow shaded region $\times$50.  \textbf{(c)} SHG spectra for trilayer. \textbf{(d)} SHG intensity compared for the same power density used on the same encapsulated sample for a laser energy of $E_L$ = 1.005~eV i.e. in resonance with IE in the bilayer, see panels (a-c), bright blue vertical line. \textbf{(e)} For a laser energy of $E_L = 0.95$ eV (indicated by vertical bright grey line in panels a and b) we plot the emission spectra in the spectral vicinity of $2E_L$ for the monolayer (black) and for the bilayer (red) using the left and right $y$-axis, respectively.  }
\end{figure*}
\textbf{Results.---} Very recently details of the rich excitonic resonances in bilayer MoS$_2$ have been uncovered \citep{Leisgang2020}. We aim to distinguish the possible SHG generation that comes from a weak, intrinsic, crystal related contribution from the possible excitonic contribution by using a tunable laser source. This allows comparing off-resonance with on-resonance experiments. \\
\indent For the experimental results shown in Fig.\ref{fig:fig1} we investigate a 2H MoS$_2$ bilayer deposited on 90 nm of SiO$_2$ on top of a Si substrate, a very typical sample configuration. We use a pulsed Ti:Sapphire laser source coupled to an optical parametric oscillator (OPO) and we scan twice the laser energy (i.e. $ 2 \times E_L $) over an energy range that corresponds to the main optical transitions in the MoS$_2$ bilayers, namely the interlayer and intralayer exciton resonances \cite{deilmann2018interlayer,slobodeniuk2019fine,gerber2019interlayer}.
\indent Experiments are performed at a temperature of $T = 4$~K in vacuum in a confocal microscope (excitation/detection spot diameter of the order of the wavelength), see Appendix. We plot in Fig.~\ref{fig:fig1}b a series of SHG spectra from the bilayer for a number of selected laser energies between $E_L=0.95$ ~eV to 1.13~eV in steps of $\approx$ 6 meV. Each SHG spectrum is a single peak that shifts in energy as we vary the excitation laser energy $E_L$. The spectral width of the SHG signal is limited by the laser pulse duration (ps). \\
\begin{figure*}
\includegraphics[width=0.98\linewidth]{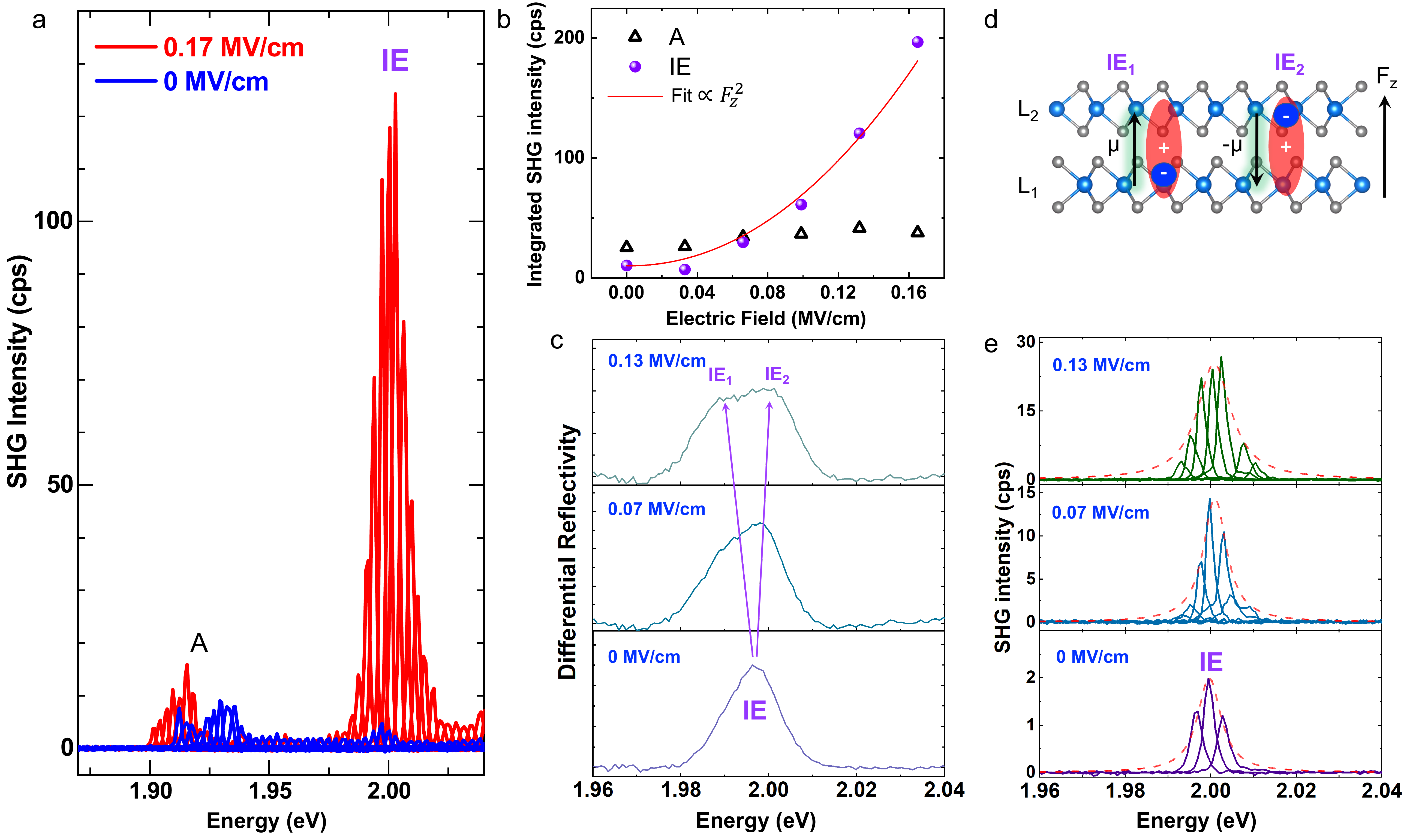}
\caption{\label{fig:fig3} \textbf{Tuning the SHG signal of a 2H MoS$_2$ bilayer in applied electric fields - Stark splitting.} Sample with contacts to apply vertical electric field $F_z$ \textbf{(a)} SHG spectra from a gated bilayer, covering intralayer A and interlayer IE exciton resonances at $F_z=0$ MV/cm (blue spectra) and $F_z=0.17$ MV/cm (red spectra) as a function of twice the laser energy $2E_L$. \textbf{(b)} comparison of SHG signal for interlayer exciton IE (solid circles) versus intralayer exciton A (open triangle) as a function of applied electric field $F_z$. The SHG amplitude of IE is fitted with a quadratic function $ \propto F_z^{2} $ (red line). \textbf{(c)} white light reflectivity for three different electric field values $F_z=0, ~0.07, ~0.13$~MV/cm. Here we plot here the first derivative of the reflectivity spectra for a better estimation of the IE energies.  \textbf{(d)} Schematic of a bilayer with 2 distinct interlayer configurations, that have opposite permanent electric dipoles  $\pm \mu$.
\textbf{(e)} SHG spectroscopy for $F_z=0, ~0.07, ~0.13$~MV/cm tuning twice the laser energy $2 \times E_L$ across the interlayer exciton resonance IE. }
\end{figure*}
\indent We start with the off-resonance case: we do not detect any SHG signal within our typical integration time (2 minutes and excitation power of a few mW) when the laser energy ($E_L$) is tuned \textit{below} half of the intralayer A:1s energy (i.e. the lowest lying direct optical transition with large oscillator strength). But surprisingly for a crystal with inversion centre, as we increase the laser energy, we see clear resonances in the SHG response, i.e. a strong variation in SHG intensity as a function of laser energy $E_L$ in Fig.\ref{fig:fig1}b. In Fig.\ref{fig:fig1}c we plot the differential white light reflectivity spectrum of this sample on the same energy scale.\\
\indent Comparing SHG spectra and white light reflectivity, we  deduce that the maxima in the SHG experiments occur when $2 \times E_L$ is in resonance with the intralayer A and  B exciton energies, separated by about 180~meV in 2H stacked MoS$_2$ bilayers \cite{paradisanos2020controlling}. For intralayer excitons the electron and the hole reside within the same layer. 
In the energy range between A and B intralayer excitons we also observe clearly that the SHG response is strongly enhanced by the interlayer exciton IE transitions \cite{deilmann2018interlayer,slobodeniuk2019fine,gerber2019interlayer}. In the absence of electric fields, there are two degenerate interlayer exciton configurations, with a hole delocalized over both layers and an electron in the top or bottom layer, respectively, as shown in Fig.~\ref{fig:fig3}d.  Interlayer excitons in homobilayer MoS$_2$ have large oscillator strength and their energy can be tuned over a wide range by the Stark effect \cite{Leisgang2020,PhysRevLett.126.037401}. We explore this tunability in experiments shown below (Fig.~\ref{fig:fig3}). \\
\indent The surprisingly strong SHG signal, which occurs in our experiments only at the excitonic resonances of the bilayer, could suggest a symmetry breaking in our non-encapsulated sample. A possible reason can be a drastically different dielectric environment experienced by top and bottom MoS$_2$ layers: For the results in Fig.~\ref{fig:fig1}b the bottom MoS$_2$ layer is on SiO$_2$ and the top layer is nominally uncovered (we perform our experiments in vacuum). To verify the impact of the dielectric environment, we have fabricated MoS$_2$ bilayer samples encapsulated in high quality hexagonal BN \cite{taniguchi2007synthesis}. This leads to a more symmetric dielectric environment \cite{stier2016probing,raja2019dielectric,rhodes2019disorder}, therefore eliminating potential sources of symmetry breaking. In addition, we also exclude the possibility of adsorbates or molecules at the sample surface in encapsulated samples which might impact the measurements as in surface SHG studies \cite{shen1989surface} and suggested in graphene \cite{doi:10.1063/1.3275740,PhysRevLett.117.123904}. In Fig.~\ref{fig:fig1}e we plot the SHG spectra as a function of twice the laser energy $2 \times E_L$ using a smaller step size of  $\approx 3$~meV between two adjacent SHG spectra. The SHG signal coming from hBN is several orders of magnitude weaker than from the MoS$_2$ bilayer, see \cite{li2013probing} and our data in the supplement. 
Also for this sample we do not detect SHG signals below 1.90~eV (intralayer A:1s resonance). As the laser energy is increased, we observe, as for the non-encapsulated sample, clear resonances. Coming back to our initial motivation, our experiments in Fig.~\ref{fig:fig1}e show that also hBN encapsulated MoS$_2$ bilayers in a far more symmetric dielectric environment show strong SHG at the exciton resonances.
These resonances are spectrally narrower in the encapsulated samples, see Fig.~\ref{fig:fig1}e, as compared to the non-encapsulated samples (Fig.~\ref{fig:fig1}b). This is consistent with TMD exciton transitions in high quality hBN encapsulated layers \cite{cadiz2017excitonic}. Comparison with white light reflectivity data in Fig.~\ref{fig:fig1}f allows attributing the different interlayer and intralayer transitions. The slight overall shift in energy of the exciton transitions for encapsulated versus non-encapsulated samples comes mainly from renormalization of all the Coulomb energies as the effective dielectric constant is different for the two bilayer samples. In addition to previously identified exciton transitions a new transition at 2.05~eV emerges in SHG spectroscopy, that we tentatively attribute to the A:2s state of the intralayer excitons. High field magneto-optics is needed to confirm this assignment \cite{goryca2019revealing} through measuring the diamagnetic shift. \\
\indent In Fig.~\ref{fig:fig1}d we plot the polarization dependence of the SHG response at the interlayer exciton resonance. We excite with linearly polarized light and the strength of the SHG signal collected in the same polarization depends on how the crystallographic axes are aligned with respect to the laser polarization. We clearly observe a 6-fold rotational symmetry expected for the space group of a 2H bilayer \cite{li2013probing}, see discussion below. This polarization dependence is a strong indication that the SHG signal is due to intrinsic effects linked to crystal and exciton symmetry. We have performed measurements at additional exciton resonances that give the same polarization dependence. {Although in principle defects with 6-fold symmetry can exist in the crystal, their direct role at the SHG at exciton resonances is unlikely to be dominant, as discussed below.} We conclude that the measured SHG signal is not due to \textit{extrinsic} effects such as impurities deposited at the surface, which can generate SHG signals unrelated to the crystal symmetry itself as studied in \textit{surface} SHG \cite{PhysRev.174.813,shen1989surface,PhysRevB.51.1425}, for graphene \cite{doi:10.1063/1.3275740,PhysRevLett.117.123904} and hBN \cite{cunha2020second}. \\
\indent We provide in Fig.~\ref{fig:fig2}a-c comparison for hBN encapsulated mono-, bi- and trilayers, for the same sample used in \cite{Leisgang2020}. Our aim is to compare the surprising SHG amplitude for bilayers with the monolayer and trilayers, for which crystal inversion symmetry is broken and as a consequence strong SHG is expected \citep{li2013probing}. Strikingly, for all the experiments on the mono-, bi- and trilayers, we see that the SHG signal is orders of magnitudes enhanced when twice the laser energy $2 \times E_L$ is in resonance with an excitonic transition, as compared with a non-resonant situation. \\
\indent Previous reports on SHG in MoS$_2$ bilayers did not focus on exciton resonances \cite{li2013probing,yu2015charge,zhao2016atomically,Klein2017,zhang2020second,Stiehm2019} and hence signals from monolayers were 3 orders of magnitude higher than for bilayers in the off-resonant case. From our measurements in panels Fig.~\ref{fig:fig2}a and b we deduce that $ E_L= $0.95~eV (i.e. SHG energy of 1.9~eV) is below the A-intralayer exciton resonance for monolayers and bilayers. 
We directly compare the measured SHG signal at this laser energy for monolayers and bilayers, see peaks in Fig.~\ref{fig:fig2}e, and find that the monolayer signal is indeed 2 orders of magnitude larger than the bilayer signal.
The situation is drastically different as we change laser energy: At $ E_L= $1.005~eV twice the laser energy $2 \times E_L$ is in resonance with the interlayer IE transition of the bilayer resulting in a strong amplification of the SHG signal while being non-resonant for a monolayer, see comparison in Fig.~\ref{fig:fig2}d. These different situations with respect to the exciton resonance result in a comparable overall SHG amplitude for mono- and bilayers. Note that Ref.~\citep{Klein2017} discusses a broad resonance with the high energy C-exciton, but no analysis of intra- versus interlayer excitons is provided, as details of these highly tunable transitions were only revealed very recently.\\
\indent The trilayer crystal has no inversion centre, as the monolayer, and SHG can be is expected for all energies. In addition, there exist several intralayer and interlayer exciton resonances \citep{Leisgang2020} that can potentially enhance SHG. For the measurements in  Fig.~\ref{fig:fig2}c we find an SHG signal with strong amplitude variations over the investigated energy range 1.87 and 2.15~eV. We identify the two intralayer transitions for excitons in the central layer (A$_{L2}$) and in the outer layers (A$_{L1}$ and A$_{L3}$) respectively, between 1.9 and 1.92~eV. Local maxima around 1.98~eV can be attributed to interlayer excitons, with B-exciton contributions around 2.1~eV. The exciton transitions for mono- and trilayers are identified by comparing with white light reflectivity \citep{Leisgang2020} shown in the supplement. \\
\indent By choosing a suitable laser energy, we have shown that the bilayer SHG signal (on-resonance) can reach the same order of magnitude as in the monolayer (off-resonance). We now show that by applying an electric field normal to the bilayer, we can further increase the SHG signal in particular at the interlayer exciton resonance. The interlayer exciton has an in-built, static electric dipole and high oscillator strength, which makes it visible in absorption in Figs.\ref{fig:fig1}c,f and \ref{fig:fig3}c. This strong absorption feature is highly tunable in energy through the Stark effect \cite{Leisgang2020}. In Fig.~\ref{fig:fig3} we show that the SHG signal is tunable in amplitude and also spectrally when an external electric field is applied to the bilayer. In previous works the effect of the electric field application on SHG response of 2H bilayers \cite{Klein2017,yu2015charge} has been addressed only in terms of the impact of doping and crystal symmetry breaking, while interlayer exciton tuning as in our resonant SHG experiment has not been observed.\\
\indent In Fig.~\ref{fig:fig3}a we compare the SHG signal of a bilayer with and without an applied electric field. For this gated sample, the IE SHG signal reaches 5 counts/s at $F_z=0$. As we apply a field of $F_z=0.17$~MV/cm this signal increases by a factor of 25 to about 125 counts/s at the IE resonance maximum. To study this tuning in more detail, we plot in Fig.~\ref{fig:fig3}b the SHG amplitude as a function of the applied electric field for six separate experiments. We see a quadratic increase of the IE SHG signal as a function of applied electric field $F_z$. This quadratic increase is due to mixing with intralayer excitons, as we show below.\\
\indent For the \textit{intralayer} exciton we record a much weaker increase of the SHG signal as $F_z$ is increased. At $F_z=0.17$~MV/cm the SHG resonance shows a slight red-shift (Fig.~\ref{fig:fig3}a), possibly indicating charging effect, i.e., a shift towards trion transitions. Charging effects as a possible source of symmetry breaking in bilayer SHG response are discussed for WSe$_2$ \cite{yu2015charge}, our data demonstrate that this has negligible effect on the SHG response at the interlayer exciton energy. \\
\indent So far we discussed the amplitude of the SHG response of the IE resonance as a function of the applied electric field $F_z$, we now turn our attention to the spectral range over which this amplification occurs. In a 2H homobilayer at $F_z=0$ two degenerate IE configurations exist: with a hole delocalized over both layers bound to an electron either in the bottom (IE$_1$) or top (IE$_2$) layer, see sketch in Fig.~\ref{fig:fig3}d. Application of a non-zero $F_z$ lifts the degeneracy of IE$_1$ and IE$_2$ as their static, permanent dipoles point in opposite directions. This leads to a Stark shift to lower and higher energy, respectively \citep{Leisgang2020}. This lifting of the degeneracy can be seen in Fig.~\ref{fig:fig3}c in white light reflectivity experiments. We extract a splitting between IE$_1$ and IE$_2$ of roughly 10~meV for $F_z=0.13~$MV/cm. In Fig.~\ref{fig:fig3}e we plot SHG spectroscopy results at the IE resonance for $F_z=0$,~0.07 and 0.13~MV/cm, respectively. In addition to the increase in SHG amplitude with $F_z$, we also observe that the spectral range for which amplification is observed is broadened. This broadening is the consequence of the energy splitting between IE$_1$ and IE$_2$ transitions in an applied electric $F_z$. The broadening can also be seen directly in Fig.~\ref{fig:fig3}a. Tuning both amplitude and spectral range of the SHG response of the bilayer is a direct consequence of exploiting the properties of interlayer excitons and will be further discussed below.

\textbf{Discussion.---} We first focus on the model considerations for SHG mediated by the interlayer excitons in bilayer MoS$_2$. Here the SHG signal is very sensitive to the electric field, being one of the most striking experimental observations. Further we briefly address other contributions to the SHG. The detailed theory is presented in the supplement.\\
\indent A TMD monolayer is described by the noncentrosymmetric point group $D_{3h}$ which lacks an inversion center and allows for the SHG. An ideal homobilayer in absence of external fields has an inversion symmetry and is described by the $D_{3d}$ point group where the SHG is forbidden. Qualitatively, this is because the constituent monolayers in the bilayer are rotated by an angle of $\pi$ with respect to each other in 2H stacking \cite{gong2013magnetoelectric}, and the electric field at a double frequency generated in one monolayer is compensated by the contribution of the other, Fig.~\ref{fig:fig1}a.\\
\indent An electric field $\bm F$ perpendicular to the monolayers breaks the symmetry and enables the SHG described by a phenomenological relation
\begin{equation}
\label{phen}
P_{2\omega,x} = \chi (E_{\omega,x}^2 - E_{\omega,y}^2), \quad P_{2\omega,y} = -2\chi E_{\omega,x}E_{\omega,y}, 
\end{equation}
where $\bm E_\omega$ is the incident field and $\bm P_{2\omega}$ is the polarization induced at a double frequency, $x$ and $y$ are the main in-plane axes of the structure. Since the bilayer is centrosymmetric, the non-linear susceptibility $\chi$ in Eq.~\eqref{phen} changes sign at the space inversion, particularly, $\chi(-F_z) = - \chi(F_z)$. Polarization dependence of the SHG predicted by Eq.~\eqref{phen} is observed experimentally, Fig.~\ref{fig:fig1}(d), showing 6-fold symmetry of the crystal structure.\\
\indent Intralayer excitons show negligible Stark shifts in applied fields \citep{Leisgang2020}, as their static dipole moment is negligible, as for excitons in separate monolayers \cite{roch2018quantum,verzhbitskiy2019suppressed}.
Therefore, due to their strong in-built electric dipole the main effect of the field $F_z$ is on the indirect excitons IE$_1$ and IE$_2$ whose degeneracy is lifted yielding $E_{{\rm IE}_i} = E_{\rm IE} \mp F_z \mu$, where $\pm \mu$ is the static dipole moment of the corresponding IE, Fig.~\ref{fig:fig3}(d). As a result of the energy splitting (Fig.~\ref{fig:fig3}c), the SHG contributions from IE$_1$ and IE$_2$ do not cancel, resulting in a strong SHG signal as the energy separation between IE$_1$ and IE$_2$ increases (Fig.~\ref{fig:fig3}a). 
The second-order susceptibility in the vicinity of the IE resonances in the linear-in-$F_z$ regime can be calculated taking into account two-photon excitation and coherent one-photon emission from the exciton states~\cite{glazov2017intrinsic} as (see supplement for details):
\begin{equation}
\label{chi:IE:field}
\chi_{\rm IE} = -2\mu F_z{\frac{|T|^2}{\Delta^2}} \frac{C_2 |\Phi_{\rm B:1s}(0)|^2 }{\left({2}\hbar\omega - E_{\rm IE} + \frac{|T|^2}{\Delta}+ \mathrm i \Gamma_{\rm IE}\right)^2}.
\end{equation}
Here $C_2$ is the combination of the interband momentum matrix element, electron charge, free electron mass and the band gap, $\Phi_{\rm B:1s}(0)$ is the envelope function of the B:1s exciton at the coinciding electron and hole coordinates, $T$ is the hole tunneling matrix element between the layers, $\Delta=E_{\rm B} - E_{\rm IE}$ is the energy splitting between the IE and B excitons (which are mixed due to the hole tunneling \cite{deilmann2018interlayer,gerber2019interlayer}), and $\Gamma_{\rm IE}$ is the damping of the IE resonance. As demonstrated in previous works~\cite{Leisgang2020,PhysRevLett.126.037401}, the IE exciton optical activity is mainly due to the mixing with B-excitons, thus, in Eq.~\eqref{chi:IE:field} the ratio $|T/\Delta|^2$ accounts for the IE-B exciton mixing in the two-photon excitation and single-photon emission channel. Equation~\eqref{chi:IE:field} holds provided $|\mu F_z| \ll \Gamma_{\rm IE}$, otherwise two peaks in the SHG spectra split by $2|\mu F_z|$ are expected to split (see supplement for details).\\
\indent Two important conclusions can be drawn from Eqs.~\eqref{phen} and ~\eqref{chi:IE:field}: First, $\bm P_{2\omega}$ is maximized when twice the laser energy is resonant with an indirect exciton state as $2 \times E_L = 2 \hbar \omega = E_{\rm IE}$. Second, the intensity of the SHG scales as $F_z^2$. Experimental SHG intensities in Fig.~\ref{fig:fig3}a,e and also in Figs.~\ref{fig:fig1}b,e,\ref{fig:fig2}b are strongly enhanced at the IE resonance. In the measurements in Fig.~\ref{fig:fig3}b the SHG amplitude grows quadratically as a function of the applied $F_z$. So both experimental observations strongly support our analysis. \\
\indent Importantly, the experiment demonstrates also the SHG on other excitonic species in bilayers, including weakly field dependent effect on intralayer A-excitons, Fig.~\ref{fig:fig3}(a,b), and also the SHG at zero field on intralayer A:$2s$ and B-excitons. A weaker effect of the electric field $\bm F$ on the A-excitons can be readily understood taking into account the fact that the A-excitons are mixed, e.g., with the interlayer B-excitons via hole tunneling. Corresponding energy distance $\Delta' = E_{\rm IE(B)} - E_{\rm A} \gg \Delta$ resulting in a weaker susceptibility to $F_z$.\\
\indent Interestingly, the SHG is also observed at the exciton resonances at $F_z=0$, i.e., in the nominally centrosymmetric situation. Although the signals are smaller than those at the resonances in non-centrosymmetric monolayer and trilayer samples, they are comparable to the non-resonant background in monolayers and trilayers, cf.~Fig.~\ref{fig:fig2}. Possible origin of SHG in this case could be the quadrupolar or magneto-dipolar SHG, where akin to graphene case~\cite{glazov2011second,Glazov2014101}, the effect is related to the light wavevector $q_z$ at the normal incidence. The symmetry analysis again yields Eq.~\eqref{phen} with $\chi \propto q_z$. Particularly, a (tiny) phase difference of $\phi \sim q_z a$ where  $a$ is the interlayer distance in the bilayer yields an imbalance of the contributions of individual monolayers and results in the resonant contributions at A and B excitons similar to calculated in~\cite{glazov2017intrinsic} for monolayers, but being by a factor of $\phi\sim 10^{-2}\ldots 10^{-3}$ smaller (see supplement). Our estimates show, however, that this effect is too small to account for the surprising experimental findings which demonstrate that the resonant susceptibilities of the bilayer are roughly one order of magnitude smaller than those of the monolayer.\\
\indent Thus, we arrive at the conclusion that, despite relatively symmetric environment of our bilayer sample, the structure lacks an inversion center. Possible options are (i) small built-in electric fields and (ii) inequivalence of the intralayer excitons in monolayers forming a bilayer. Option (i) results in the replacement of $F_z$ by $F_z+F_0$ where $F_0$ is the normal component of the build-in field. In this case at $F_z=-F_0$ the effectively symmetric situation can be realized. 
While the measured dependence of the SHG intensity for IE, Fig.~\ref{fig:fig3}, does not contradict this scenario, note the minimum at $F_z \approx -0.035$~MV/cm, the A:1$s$ intensity does not significantly drop in this region. 
Alternatively, we may suppose that the build-in field is inhomogeneous in the sample plane within our detection spot and the contributions of the A:1$s$ and IE states come from slightly different nanoscopic regions. \\
\indent Option (ii) implies that the energies of A- and B-excitons or their oscillator strengths are inequivalent in the two constituent monolayers that form the bilayer. This may be due to the dielectric disorder, which, although suppressed, can still be present in state-of-the-art, hBN encapsulated samples~\cite{raja2019dielectric}. In this situation the IE is mainly activated by the electric field, while intralayer excitons are less sensitive to $F_z$ (see Supplement).  Inhomogeneous broadening effects due to disorder, impurities, etc. are inevitable~\cite{raja2019dielectric}, in particular in MoS$_2$ samples with different type of defects reaching densities of 10$^{10}$ to 10$^{12}$ per cm$^{-2}$ \cite{rhodes2019disorder}. But the crucial parameter is the ratio between the size of the inhomogeneity, $l$ and the laser spot diameter, $\delta$. 
For $l\ll\delta$, inhomogeneous broadening effects will effectively cancel out and the structure will effectively maintain the space inversion. 
However, for $l\gtrsim \delta$, the inhomogeneities will lift the inversion symmetry, thus a contribution of one monolayer will dominate over the other and SHG occurs. Sizeable effects due to disorder require a comparable or larger spatial scale than the excitation spot. Inhomogeneities on the order of one micrometer can occur even in hBN encapsulated samples~\cite{raja2019dielectric}. It is important to note that sample inhomogeneity does not always translate in a large shift of the excitonic transition energy, as bandgap shifts and exciton binding energy changes partially compensate~\cite{raja2019dielectric}.
We stress that the six-fold symmetry of the SHG signal (see Supplement) clearly rules out, e.g., an in-plane asymmetry due to the strain or in-plane fields \citep{mennel2018optical}.  \\

\textbf{Conclusion}.--- 
In summary, we show strong and electrically tunable exciton-mediated SHG in 2H MoS$_2$ bilayers that can surpass the off-resonance monolayer signal. Drastic enhancement of the SHG amplitude is observed when twice the laser energy is in resonance with the excitonic transitions. At the interlayer exciton resonance, we tune the SHG signal by over an order of magnitude in  electric fields applied perpendicular to the layer and demonstrate that the spectral width of the SHG resonance increases. We relate the SHG in applied electric field to the Stark splitting of the interlayer exciton and its mixing with intralayer excitons. Interlayer excitons form due to coupling between two layers. This coupling depends on several parameters such as layer distance and twist angle. Changing these parameters therefore allows tuning of SHG in van der Waals stacks. We identify several possible mechanisms for the strong zero-field SHG signal, such as small in-built electric fields in the structure. Disorder, possibly related to defects in the layer or in the dielectric environment, could result in non-equivalence of the intralayer excitons in the different layers within the optical detection spot on our sample structure. Our scheme for tuning SHG can be applied to a variety of other systems, such as  homobilayer MoSe$_2$ with strong interlayer exciton resonances \cite{PhysRevB.97.241404},  heterobilayers which host interlayer excitons with high oscillator strength   for example MoSe$_2$/WSe$_2$ with hybridized conduction states  \cite{alexeev2019resonantly} and WSe$_2$/WS$_2$ with hybridized valence states  \cite{tang2021tuning}. The impact of concentration and type of defects states on the non-linear optical properties merits further investigation \cite{refaely2018defect,rhodes2019disorder}.
\section{Methods}
\textbf{Sample fabrication.-}  90~nm thick SiO$_2$/Si substrates were cleaned for 10~minutes in aceton and isopropanol using an ultrasonication bath and were subsequently exposed in oxygen plasma for 60~seconds. Bulk MoS$_2$ (2D Semiconductors) was first exfoliated on Nitto Denko tape and the exfoliated areas were attached on a polydimethylsiloxane (PDMS) stamp, supported by a microscope glass slide. Monolayers, 2H-bilayers and trilayers were identified based on the optical contrast under an optical microscope prior to transfer on the SiO$_2$/Si substrate. For the hBN encapsulated samples, hBN flakes were first exfoliated on a Nitto Denko tape from high quality bulk crystals~\cite{taniguchi2007synthesis} while the same PDMS-assisted transfer process on SiO$_2$/Si substrates was followed. A staircase sample of monolayers, 2H-bilayers and trilayers was susbequently transferred and capped in hBN. Between every transfer step, annealing at 150$^{o}$C was applied for 60~min. For the electric field device, the same process was followed including the additional transfer of few-layered graphite (FLG) flakes. FLG flakes were aligned according to Au contacts on a pre-patterned substrate to act as electrodes for the electric field measurements. The precise sequence of the complete stack from bottom to top includes hBN/FLG/hBN/2H-MoS$_2$/hBN/FLG.\\
\indent \textbf{Optical spectroscopy Set-up.-}  Optical spectroscopy is performed in a home-built micro-spectroscopy set-up assembled around a closed-cycle, low vibration helium cryostat with a temperature controller ($T=4$~K to 300~K). For SHG measurements we use ps pulses, generated by a tunable optical parametric oscillator (OPO) synchronously pumped by a mode-locked Ti:sapphire laser as in \cite{wang2015giant}. SHG signal is collected in reflection geometry. For low temperature white light reflectance measurements a white light source; a halogen lamp is used with a stabilized power, focused initially on a pin-hole that is imaged on the sample. 
The emitted and/or reflected light was dispersed in a spectrometer and detected by a Si-CCD camera. The excitation/detection spot diameter is 1~$\mu m$, i.e., smaller than the typical size of the homobilayers. We obtained differential reflectivity from reflectivity spectra as $(R_\text{ML}-R_\text{sub})/R_\text{sub}$, where $R_{\rm ML}$ is the intensity reflection coefficient of the sample with the MoS$_2$ layer and $R_{\rm sub}$ is the reflection coefficient of the hBN/SiO$_2$ stack.\\ 

\section{Supplement}
\subsection{Sample Details}
\begin{figure*}
\includegraphics[width=0.79\linewidth]{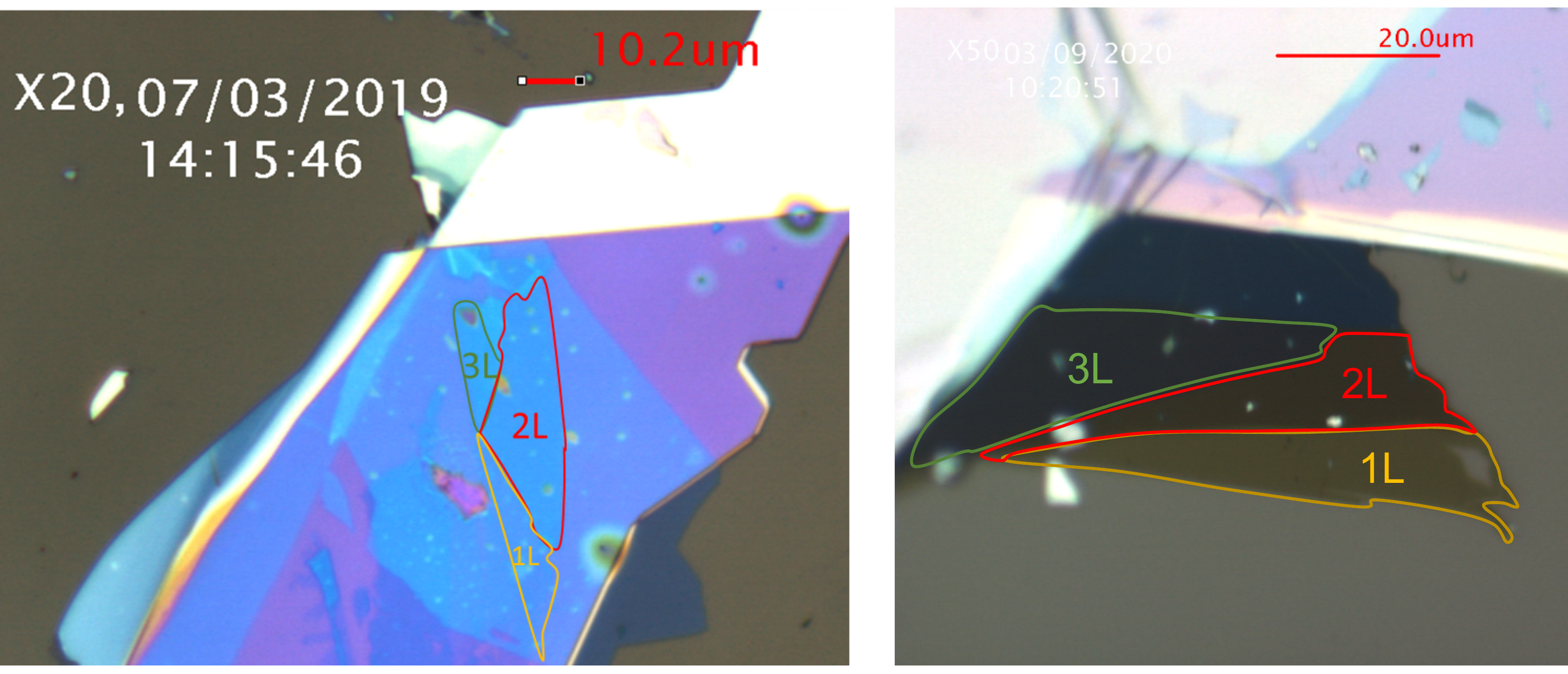}
\caption{\textbf{Optical images of hBN encapsulated and bare MoS$_2$ on SiO$_2$.}}
\label{figS1}
\end{figure*}

Van der Waals heterostructures (vdWHs) were fabricated by stacking two-dimensional materials via a dry-transfer technique. 2H-MoS$_2$ crystals (2D Semiconductors) and synthetic hBN \cite{taniguchi2007synthesis} were subjected to micromechanical cleavage on Nitto Denko tape, then exfoliated again on a polydimethylsiloxane (PDMS) stamp placed on a glass slide for optical inspection. Sequential deterministic stamping of the selected flakes was then applied to build the complete stack. Optical images of hBN encapsulated but also bare MoS$_2$ in SiO$_2$  are shown in Fig.~\ref{figS1}. The thickness of the bottom hBN layers was selected to optimize the oscillator strength of the interlayer exciton (IE) in bilayer MoS$_2$ \cite{robert2018optical}.

\subsection{Experimental setup}
\begin{figure}[h]
\centering
\includegraphics[width=0.99\linewidth]{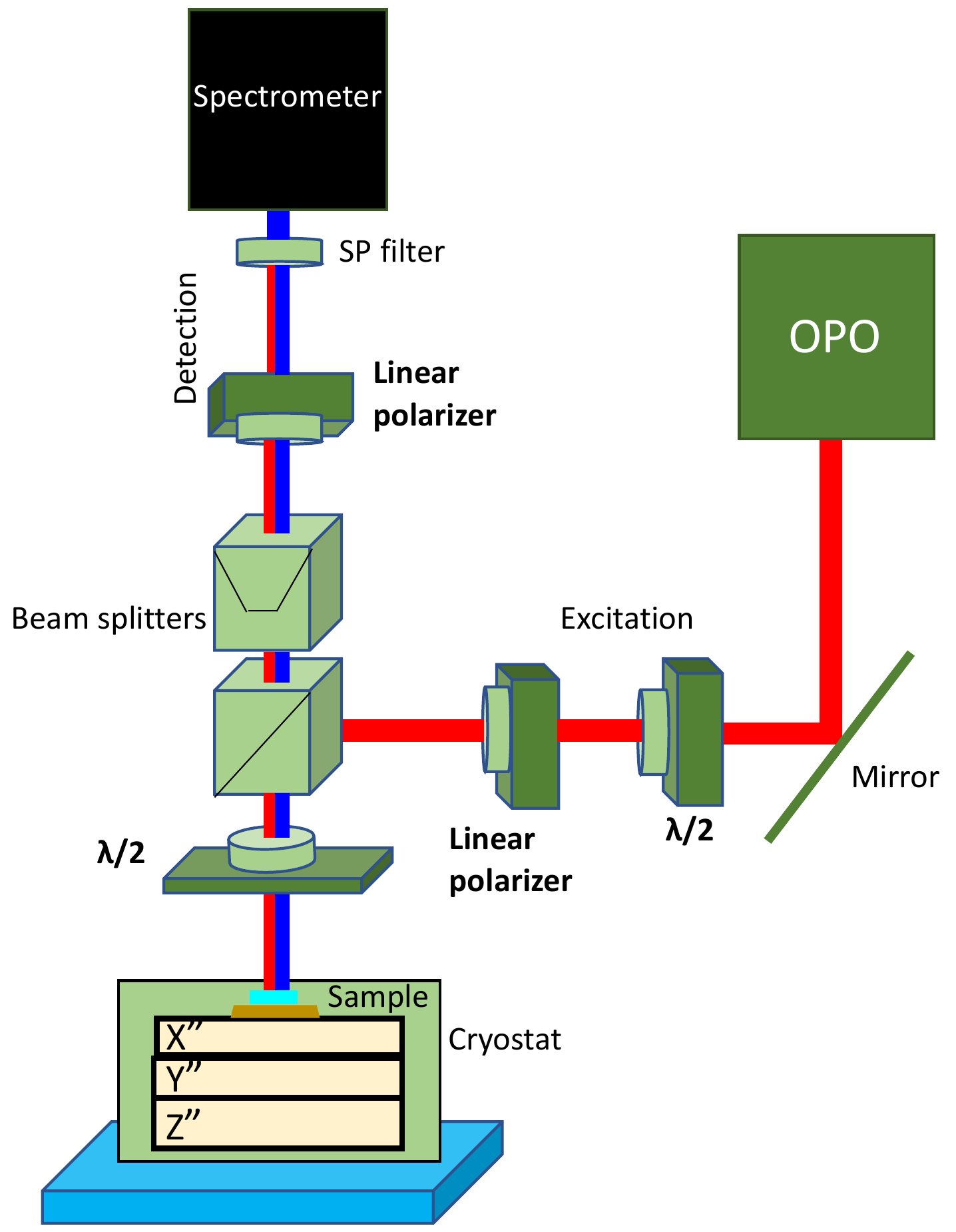}\caption{\textbf{Experimental setup used for the collection of the SHG spectra.}}
\label{figS2}
\end{figure}

The SHG spectra were recorded with the setup sketched in Fig.~\ref{figS2}. An Optical Parametric Oscillator (OPO) is aligned to a closed-cycle cryostat to excite the sample under investigation. A combination of linear polarizers and halfwave plates allows the control of excitation and detection polarization for the polarization-resolved measurements. The light is focused onto the sample at 4.2~K using a microscope objective (NA=0.8). The position of the sample with respect to the focus can be adjusted with cryogenic nanopositionners. The reflected light from the sample is sent to a spectrometer with a 150 grooves per millimeter grating. The spectra are recorded by a liquid-nitrogen cooled charged coupled device (CCD) array. 
Low temperature reflectance measurements were performed using a halogen lamp as a white-light source with a stabilized power supply focused initially on a pin-hole that is imaged on the sample. The reflected light was dispersed in a spectrometer and detected by the same Si-CCD camera. The excitation/detection spot diameter is $\approx1\mu$m, i.e. smaller than the typical diameter of the sample. 

\subsection{Assignment of resonances in monolayers and trilayers}
\begin{figure*}
\includegraphics[width=0.69\linewidth]{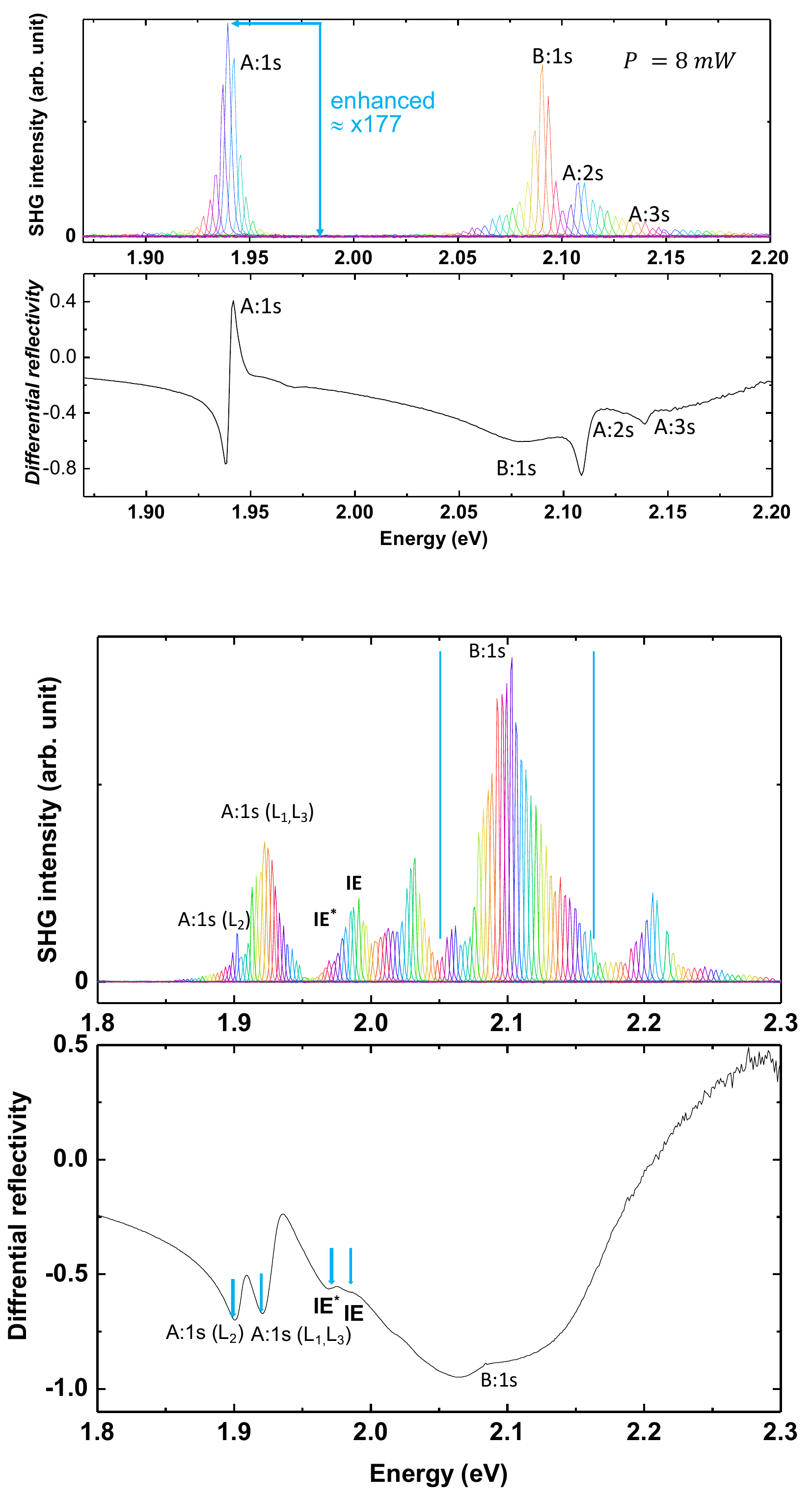}
\caption{\textbf{Comparison between SHG and differential reflectivity spectra of monolayer and trilayer MoS$_2$.}}
\label{figS3}
\end{figure*}

Monolayers and trilayers have also been investigated via SHG spectroscopy. For the assignment of the different excitonic resonances in the SHG spectra, differential reflectance has also been collected from the same sample areas  (compare top and bottom spectra in Fig.~\ref{figS3}). Besides 1s states, excited states of the A exciton can be observed in SHG and verified by reflectivity. Note that interlayer exciton states are clear in the SHG spectra of trilayers, see also \citep{Leisgang2020}.  

\subsection{Background SHG}
\begin{figure*}
\centering
\includegraphics[width=0.69\linewidth]{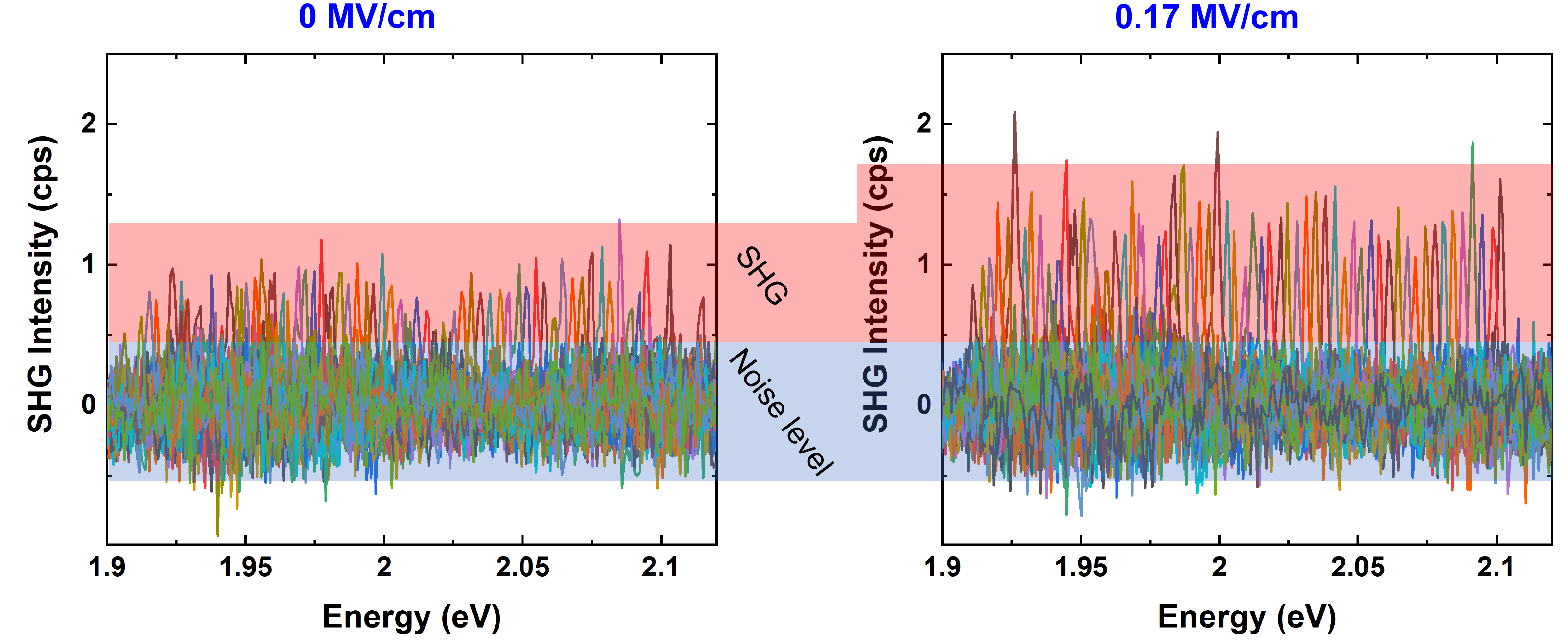}\caption{\textbf{Background SHG contribution from the device excluding MoS$_2$ at zero and finite electric field.}}
\label{figS4}
\end{figure*}

For the gated bilayer device it is important to identify the background contribution in the SHG. Few-layered graphene (FLG), as well as hBN can provide a contribution in the SHG spectra. In Fig.~\ref{figS4} we show the noise level of the CCD camera as well as the background SHG originating from hBN and FLG under zero and finite electric fields. In this measurement, we collected the SHG coming from the same stack of the gated sample, but without MoS$_2$. The background SHG is weak ($<2$counts/second) compared to the signal from MoS$_2$ and it is wavelength independent. This unambiguously confirms the spectral signature of the excitonic resonances in SHG of MoS$_2$ bilayers.

\subsection{Polarization-resolved SHG}
\begin{figure}[h]
\centering
\includegraphics[width=85mm]{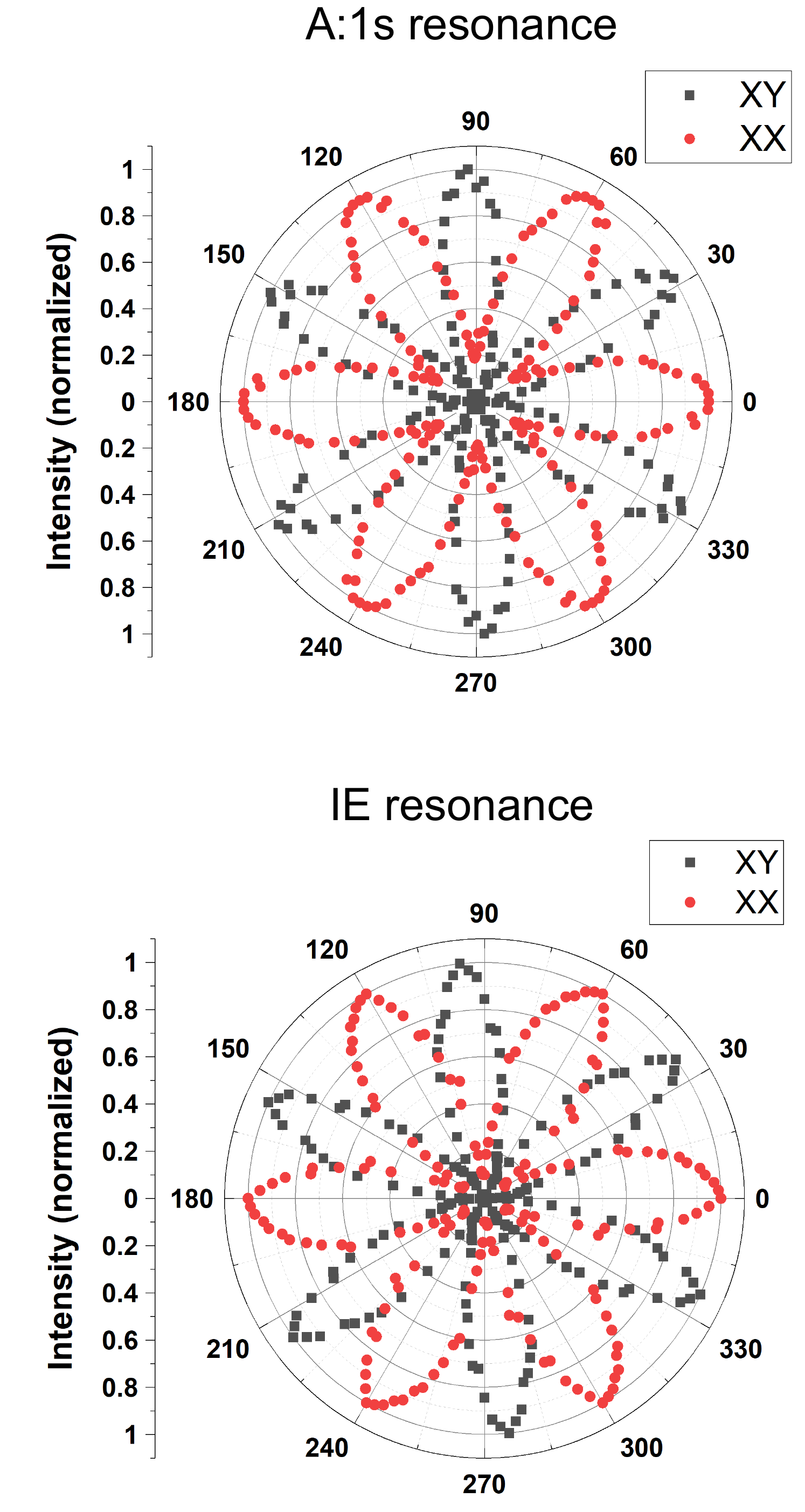}\caption{\textbf{Comparison between SHG and differential reflectivity spectra of monolayer and trilayer MoS$_2$.}}
\label{figS5}
\end{figure}

Polarization-resolved SHG has been applied to verify the symmetry and crystallographic orientation of MoS$_2$ bilayers \cite{li2013probing,PhysRevB.87.201401}. The intensity of the SHG as a function of the polarization angle of the fundamental laser is shown in Fig.~\ref{figS5}. Notice that similar angle dependence is acquired when the excitation is in resonance with the A:1s and IE states. This SHG intensity is dependent on the crystallographic orientation of the sample. The symmetric 6-fold co- (XX) and cross-linear (XY) polarization over 2$\pi$ reveals that any contribution from strain effects and low symmetry defects is negligible \cite{cunha2020second,mennel2018optical}.

\subsection{$\mbox{hBN}$ nonlinear contribution}
\begin{figure}
\centering
\includegraphics[width=85mm]{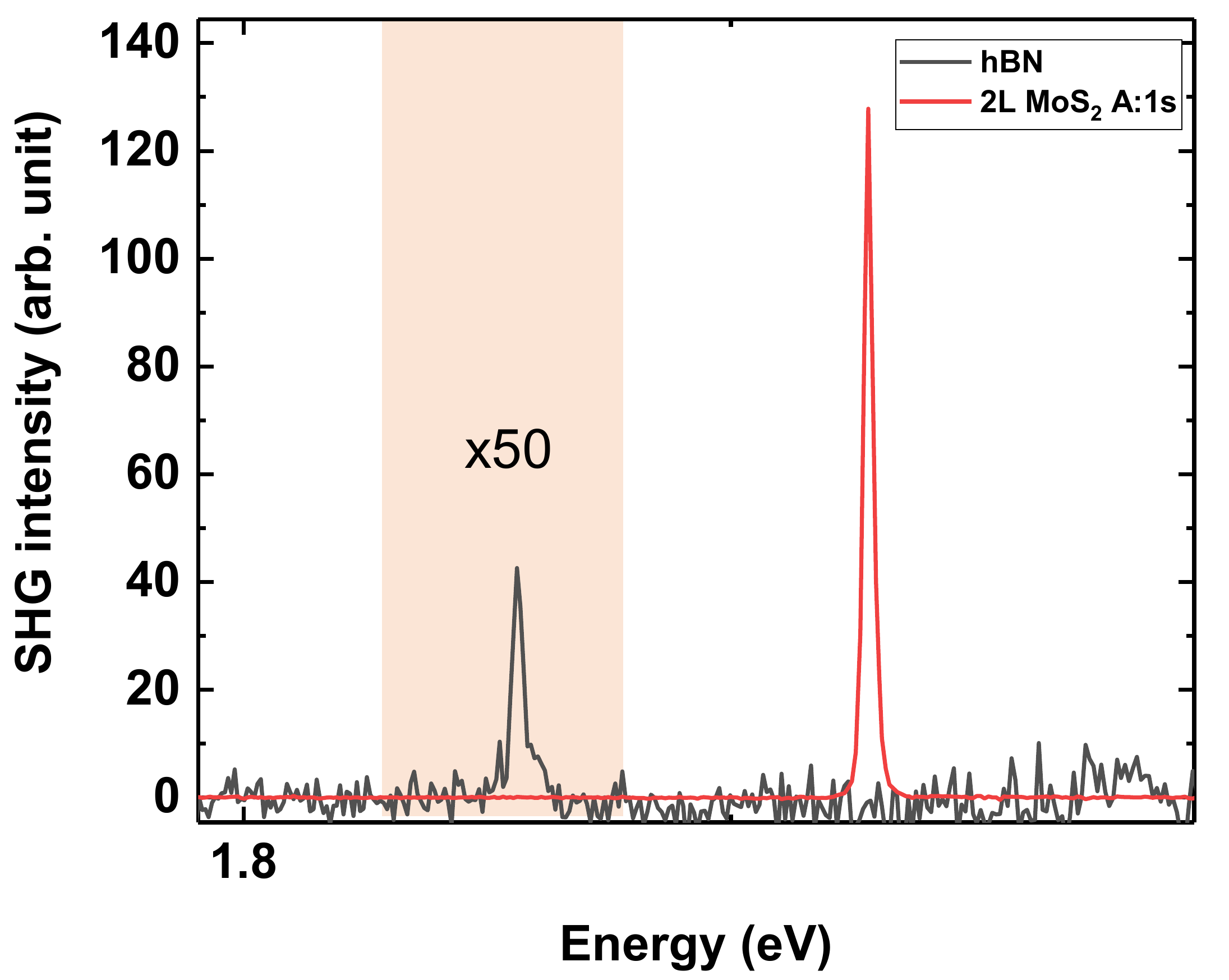}\caption{\textbf{Comparison between SHG of bilayer MoS$_2$ with hBN. Notice that the SHG from hBN is multiplied by a factor of 50.}}
\label{figS6}
\end{figure}

In the encapsulated bilayers with hBN, we measure the SHG strength of the surrounding hBN in the total signal. In Fig.~\ref{figS6} we compare the SHG intensity coming from the surrounding hBN with the MoS$_2$ bilayer when the excitation is kept constant and tuned into resonance with the A:1s state. The SHG coming from MoS$_2$ is two orders of magnitude stronger.

\subsection{Validation of the nonlinear processes}
\begin{figure}
\centering
\includegraphics[width=85mm]{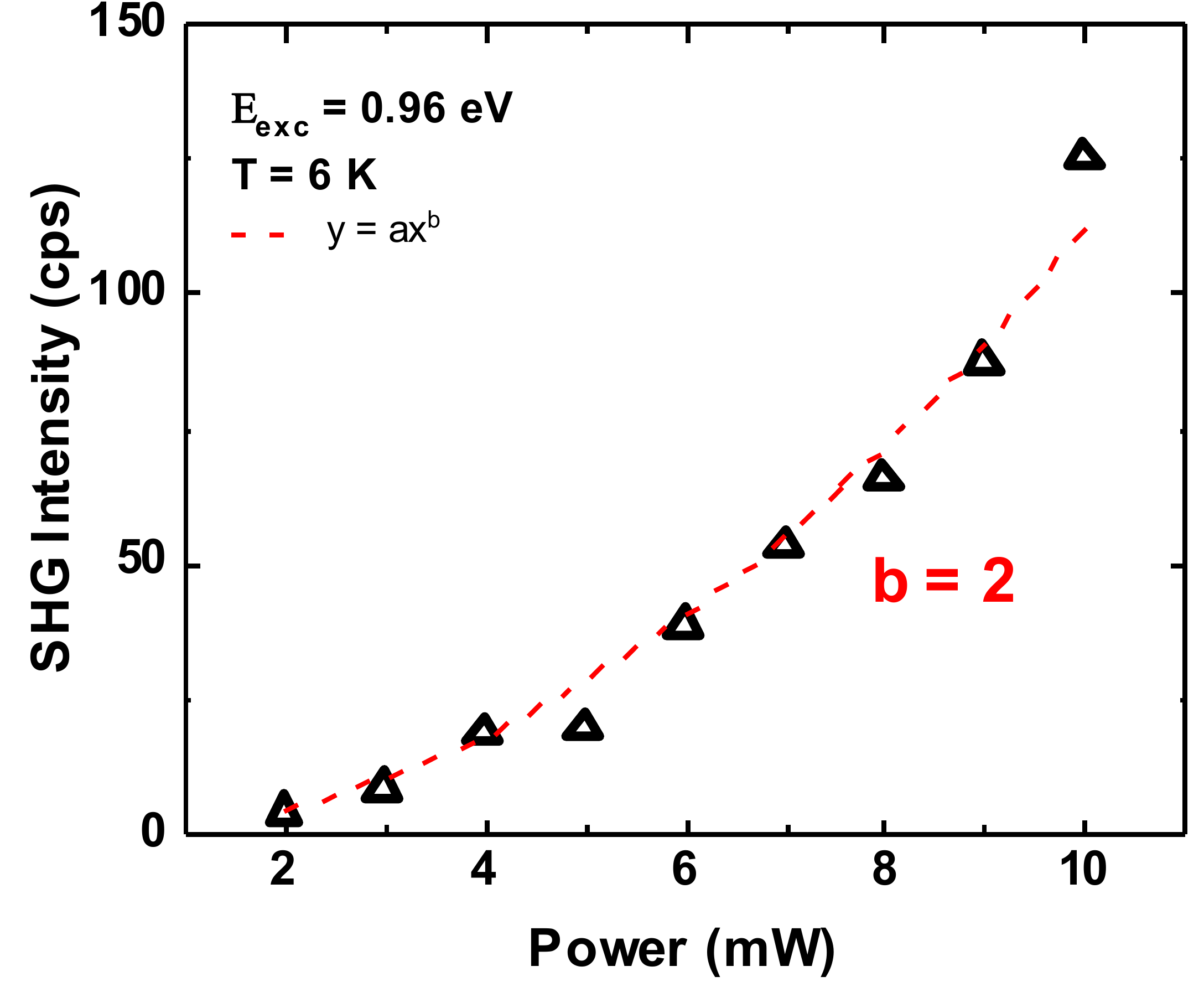}\caption{\textbf{Power dependent SHG in resonance with the A-exciton clearly demonstrates a two-photon process.}}
\label{figS7}
\end{figure}

Power dependent measurements have been performed in MoS$_2$ bilayers. In Fig.~\ref{figS7} the SHG intensity is plotted as a function of laser power. The excitation energy is tuned in resonance with the A:1s state. The quadratic power law confirms the two-photon process.


\section{Theory}

This section contains the phenomenological and microscopic theory describing the SHG in bilayer structures.

\subsection{SHG in an ideal centrosymmetric bilayer}

Bilayer has a $D_{3d}$ point symmetry with an inversion center. For the light propagating along the normal $z$ there is a contribution caused by the radiation wavevector $q_z$ (quadrupolar contribution) which results from the inhomogeneity of the incident and emitted electric field distribution:
\begin{equation}
\label{phen:2ML}
P_{2\omega,x} = q_z\chi' (E_{\omega,x}^2 - E_{\omega,y}^2), \quad P_{2\omega,y} = -2q_z\chi'  E_{\omega,x}E_{\omega,y}.
\end{equation}

To illustrate the effect let us study the electrodynamical mechanism leading to Eq.~\eqref{phen:2ML}. For this mechanism we consider both monolayers forming a bilayer as electronically decoupled, the distance between the layers is $a$. Disregarding multiple reflections of the second harmonics radiation from both monolayers we obtain the reflected and transmitted amplitudes of the second harmonics in the form
\begin{subequations}
\label{electrodyn}
\begin{align}
E_{2\omega}^r = S\chi E_\omega^2 (1-\text{e}^{i(q_{2\omega}+2q_\omega)a}),\\
\qquad
E_{2\omega}^t = S\chi E_\omega^2 (\text{e}^{iq_{2\omega}a}-\text{e}^{i2q_\omega a}).
\end{align}
\end{subequations}
Here $S$ is the coefficient describing the propagation of the SHG field from the sample, $\chi$ is the second-order susceptibility of a monolayer~\cite{glazov2017intrinsic}, and
$q_\Omega = {\Omega n_\Omega / c}$ with $n_\Omega$ being the refraction index at the frequency $\Omega$. It is noteworthy that for a bilayer the products $q_\omega a, q_{2\omega} a \ll 1$. As a result, the intensity of the SHG from the bilayer is suppressed as compared with the monolayer by a factor of $\sim q_{2\omega} a \ll 1$, resulting in the small SHG signal from the bilayer (see the main text for discussion). Accounting for the electronic coupling between the layers may result in additional contributions to the SHG which has the same smallness $\propto q_{\omega} a$.

In Eqs.~\eqref{electrodyn} we neglected the differences between the transmission coefficients of the monolayers and unity at both fundamental and double frequencies and disregarded the effect of the substrate on the fields. While the transmission coefficient of the monolayer $t_\omega$ at a fundamental frequency is very close to $1$ because the incident radiation is non-resonant, the effect of the monolayers on the transmission and reflection of the second harmonic can be sizeable due to the exciton resonances. With account for multiple reflections of the second harmonics from the monolayers, we have an enhancement of both the reflected and transmitted waves. In the reflection geometry the enhancement factor [cf. Ref.~\cite{PhysRevLett.123.067401}]
\begin{equation}
\label{Fr}
F_r=
{1+t_{2\omega} \frac{r_{2\omega} e^{i2q_{2\omega} a}}{1-r_{2\omega}^2 e^{i2q_{2\omega} a}} \approx \frac{1}{1-r_{2\omega}}},
\end{equation}
where $r_{2\omega}$ is the reflection coefficient from the monolayer at the frequency $2\omega$, $t_{2\omega} = 1+r_{2\omega}$, and in the last equation we neglected the phase acquired by the field propagating between the layers. Making use of the explicit form of the monolayer reflection coefficient we recast $F_r$ as
\[
{F_r = \frac{\omega_0 - 2\omega - i(\Gamma_0+\Gamma)}{\omega_0 - 2\omega - i(2\Gamma_0 + \Gamma)},}
\]
where $\omega_0$ is the exciton resonance frequency, $\Gamma_0$ and $\Gamma$ are its radiative and non-radiative damping.

The expression for the enhancement factor in the transmission geometry has somewhat more complex form
\begin{multline}
\label{Ft}
F_t =\\
 \frac{ \text{e}^{iq_{2\omega}a} {t_{2\omega}\over 1-r_{2\omega}^2 \text{e}^{2iq_{2\omega}a}}
-\text{e}^{i2q_\omega a} \left(1 
+ {t_{2\omega} r_{2\omega} \text{e}^{2iq_{2\omega}a} \over 1-r_{2\omega}^2 \text{e}^{2iq_{2\omega}a}}\right)}{t_{2\omega}\text{e}^{iq_{2\omega}a}-\text{e}^{i2q_\omega a}}.
\end{multline}
{It follows from Eqs.~\eqref{Fr} and \eqref{Ft} that in the vicinity of the exciton resonance the enhancement due to multiple reflections is unimportant.}

It is instructive to discuss the relation between the effects described above and the surface SHG effect both at oblique and normal incidence studied in a number of works, including Refs.~\cite{PhysRev.174.813,shen1989surface,PhysRevB.33.8254,PhysRevB.35.1129,PhysRevB.51.1425}. It has been previously established that either the surface of a media has a reduced symmetry, e.g.,~\cite{PhysRevB.51.1425} and allows for the SHG by itself, or the SHG arises due to the quadrupolar or magneto-dipolar transitions ($qE^2$ or $EB$ mechanisms), e.g., ~\cite{PhysRev.174.813}. The mechanism discussed above belongs to the second type of the effects.

To conclude this part, in  ideal centrosymmetric bilayer the SHG is possible due to the effects related to the finite wavevector of the radiation and a finite width of the bilayer. Because of this, the effect is suppressed in pristine bilayer as compared to the non-centrosymmetric monolayer.

\subsection{Electric field induced mechanism}\label{sec:field}

\subsubsection{Minimal model}

Let us first consider the minimal model demonstrating the electric field induced SHG in bilayers. To that end motivated by microscopic analysis presented in Refs.~\cite{Leisgang2020,PhysRevLett.126.037401} we consider the model of two interlayer exciton states IE$_{1,2}$ and two intralayer exciton states B$_{1,2}$. These four states are close in energy and are effectively mixed by the hole tunneling. Let $T$ be the (hole) tunneling constant which mixes corresponding IE and B exciton states. The Hamiltonian of the system in the presence of the electric field $F_z$ reads (the basic states are
 IE$_1$, B$_1$, IE$_2$, B$_2$):
\begin{equation}
\label{Ham}
\mathcal H = \begin{pmatrix}
E_{\rm IE} - \mu F_z & T &0 &0\\
T^* & E_{\rm B} & 0 & 0\\
0 & 0 & E_{\rm IE} + \mu F_z & T\\
0 & 0 & T^* & E_{\rm B}
\end{pmatrix}.
\end{equation}
Here $\mu$ is the indirect exciton static dipole matrix element along the structure normal (see main text for details).

{For each pair of levels IE$_1$ \& B$_1$ and IE$_2$ \& B$_2$ we have a $2\times 2$ Hamiltonian which is readily diagonalized as 
\begin{subequations}
\label{states:2:exact}
\begin{align}
 E_{1,\pm}(F_z) &= \frac{E_{\rm IE} - \mu F_z + E_{\rm B}}{2} \\
& \pm \sqrt{\left(\frac{E_{\rm IE} - \mu F_z - E_{\rm B}}{2}\right)^2 + |T|^2}, \nonumber\\
 E_{2,\pm}(F_z) &= E_{1,\pm}(-F_z),
\end{align}
with the eigenfunctions
\begin{align}
 |1,\pm\rangle &= \alpha_\pm(F_z) |\mathrm{IE}_1\rangle + \beta_\pm(F_z) |\mathrm B_1\rangle,\\
 |2,\pm\rangle &= \alpha_\pm(-F_z) |\mathrm{IE}_2\rangle + \beta_\pm(-F_z) |\mathrm B_2\rangle.
\end{align}
The coefficients $\alpha_\pm(F_z)$, $\beta_\pm(F_z)$ read
\begin{align}
\label{alpha:beta}
\alpha_\pm(F_z) &= \frac{E_{1,\pm}-E_{\mathrm B}}{T^*} \beta_\pm(F_z), \\
 \beta_\pm(F_z) &= \frac{1}{\sqrt{1+ \frac{(E_{1,\pm}-E_{\mathrm B})^2}{|T|^2}}} \nonumber.
\end{align}
\end{subequations}
}

In order to derive compact analytical expressions we assume that the distance between the IE and B excitons $\Delta = E_B - E_{\rm IE}$ exceeds by far both $|T|$ and $|\mu F_z|$, and also that $|\mu F_z| \ll |T|$. Hence, the {eigenenergies and eigenstates are}
\begin{widetext}
\begin{subequations}
\label{states}
\begin{align}
&\widetilde{{\rm IE}_1}: \quad E_{\widetilde{{\rm IE}_1}} \approx E_{\rm IE} - \mu F_z - \frac{|T|^2}{\Delta}, \quad | \widetilde{{\rm IE}_1}\rangle = |{\rm IE}_1\rangle - {\frac{T^*}{\Delta}\left( 1-\frac{\mu F_z}{\Delta}\right) |\mathrm B_1\rangle},\\
&\widetilde{{\rm IE}_2}: \quad E_{\widetilde{{\rm IE}_2}} \approx E_{\rm IE} + \mu F_z - \frac{|T|^2}{\Delta}, \quad | \widetilde{{\rm IE}_2}\rangle = |{\rm IE}_2\rangle - {\frac{T^*}{\Delta}\left( 1+\frac{\mu F_z}{\Delta}\right) |\mathrm B_2\rangle},\\
&\widetilde{{\rm B}_1}: \quad E_{\widetilde{{\rm B}_1}} \approx E_{\rm B} + \frac{|T|^2}{\Delta}\left( 1-\frac{\mu F_z}{\Delta}\right), \\
 & \quad\quad\quad | \widetilde{{\rm B}_1}\rangle = {\left[1-\frac{|T|^2}{2\Delta^2}\left(1-\frac{2\mu F_z}{\Delta}\right)\right]|\mathrm B_1\rangle + \frac{T}{\Delta}\left( 1-\frac{\mu F_z}{\Delta}\right) |\mathrm{IE}_1\rangle}, \nonumber\\
&\widetilde{{\rm B}_2}: \quad E_{\widetilde{{\rm B}_2}} \approx E_{\rm B} + \frac{|T|^2}{\Delta}\left( 1+\frac{\mu F_z}{\Delta}\right),  \\
 & \quad\quad\quad | \widetilde{{\rm B}_2}\rangle = {\left[1-\frac{|T|^2}{2\Delta^2}\left(1+\frac{2\mu F_z}{\Delta}\right)\right]|\mathrm B_2\rangle + \frac{T}{\Delta}\left( 1+\frac{\mu F_z}{\Delta}\right) |\mathrm{IE}_2\rangle} \nonumber.
\end{align}
\end{subequations}
\end{widetext}

Let $M_2$ and $M_1$ be the two-photon and the one-photon excitation matrix elements of the B-exciton.
The susceptibilities of B$_{1,2}$ excitons in isolated monolayers $T=0$ read ($C$ is a constant)~\cite{glazov2017intrinsic}
\begin{equation}
\label{suscept:B}
\chi_{\mathrm B, 1,2} = \pm C\frac{M_1 M_2}{{2}\hbar\omega - E_{\mathrm B} + \mathrm i \Gamma_{\rm B}}.
\end{equation}
For a bilayer at $F_z=0$ the sum $\chi_{\mathrm B,1}+ \chi_{\mathrm B,2}=0$ signifying the presence of an inversion center.
 
The SHG generation results from the mixing of those states with B-excitons. {Generally, the susceptibility can be written as
\begin{equation}
\label{suscept:alpha:beta}
\chi =- C M_1 M_2 \sum_{j=1}^2\sum_\pm (-1)^{j}\frac{|\beta_\pm(F_z)|^2}{2\hbar\omega - E_{j,_\pm}(F_z) + \mathrm i \Gamma_{j,_\pm}}. 
\end{equation}
}
As a result, {for the weak tunneling and field} we obtain from Eqs.~\eqref{states}
\begin{widetext}
\begin{subequations}
\label{suscept}
\begin{align}
&\widetilde{{\rm IE}_1}: \quad \chi_{\mathrm{IE}_1} = {\frac{|T|^2}{\Delta^2}\left( 1-\frac{2\mu F_z}{\Delta}\right)  }\frac{CM_1M_2}{{2}\hbar\omega - E_{\rm IE} + \mu F_z {+} \frac{|T|^2}{\Delta}+ \mathrm i \Gamma_{\rm IE}},\\
&\widetilde{{\rm IE}_2}: \quad \chi_{\mathrm{IE}_2} = -{\frac{|T|^2}{\Delta^2}\left( 1+\frac{2\mu F_z}{\Delta}\right)  } \frac{CM_1M_2}{{2}\hbar\omega - E_{\rm IE} - \mu F_z {+} \frac{|T|^2}{\Delta}+ \mathrm i \Gamma_{\rm IE}},\\
&\widetilde{{\rm B}_1}: \quad \chi_{\mathrm B_1} =  {\left[1-\frac{|T|^2}{2\Delta^2}\left(1-\frac{2\mu F_z}{\Delta}\right)\right]}\frac{CM_1 M_2}{{2}\hbar\omega - E_{\mathrm B}-\frac{|T|^2}{\Delta}\left( 1-\frac{\mu F_z}{\Delta}\right)+ \mathrm i \Gamma_{\rm B}},\\
&\widetilde{{\rm B}_2}: \quad \chi_{\mathrm B_2} = - {\left[1-\frac{|T|^2}{2\Delta^2}\left(1+\frac{2\mu F_z}{\Delta}\right)\right]}\frac{CM_1 M_2}{{2}\hbar\omega - E_{\mathrm B}-\frac{|T|^2}{\Delta}\left( 1+\frac{\mu F_z}{\Delta}\right)+ \mathrm i \Gamma_{\rm B}}.
\end{align}
\end{subequations}
\end{widetext}
The total susceptibility
\begin{equation}
\label{chi:tot}
\chi = \chi_{\rm IE_1} + \chi_{\rm IE_2} + \chi_{\rm B_1} + \chi_{\rm B_2}.
\end{equation}
It vanishes at $F_z=0$. {At $F_z \ne 0$ the inversion symmetry is broken and effect arises both due to the energy shifts of the excitons and due to the linear in $F_z$ contributions to the oscillator strength.}

To get better insight into the SHG in bilayer materials and for the comparison with available experiments it is instructive to analyze the situation where the electric field is so weak such that $|\mu F_z| \ll \Gamma_{\rm B, IE,\ldots}$. The susceptibility in the vicinity of the IE excitons takes the form 
\begin{multline}
\label{chi:IE}
\chi_{\rm IE} = \chi_{\rm IE_1} +  \chi_{\rm IE_2} \\
= -\frac{2\mu F_z}{{2}\hbar\omega - E_{\rm IE} + \frac{|T|^2}{\Delta}+ \mathrm i \Gamma_{\rm IE}} {\frac{CM_1M_2 |T|^2}{\Delta^2}}\times \\
\left(\frac{2}{\Delta} + \frac{1}{{2}\hbar\omega - E_{\rm IE} + \frac{|T|^2}{\Delta}+ \mathrm i \Gamma_{\rm IE}}\right).
\end{multline}
For reasonable $\Delta > \Gamma_{\rm IE}$ the second term in parentheses dominates in the vicinity of the resonance. It results in Eq. (2) of the main text with the notation $C_2 |\Phi_{\mathrm B:1s}(0)|^2 = C M_1 M_2$, cf. Ref.~\cite{glazov2017intrinsic}. Similar expression can be derived describing the susceptibility in the vicinity of the B excitons:
\begin{multline}
\label{chi:B}
\chi_{\rm B} = \chi_{\rm B_1} +  \chi_{\rm B_2}\\
 = -\frac{2\mu F_z}{{2}\hbar\omega - E_{\rm B} - \frac{|T|^2}{\Delta}+ \mathrm i \Gamma_{\rm B}} {\frac{CM_1M_2 |T|^2}{\Delta^2}}\times \\
\left(\frac{1}{\Delta} - \frac{1}{{2}\hbar\omega - E_{\rm B} - \frac{|T|^2}{\Delta}+ \mathrm i \Gamma_{\rm B}}\right).
\end{multline}

\subsubsection{Extensions of the model}

The deficiency of the model above is the absence of the effect on A-excitons and also the neglect of the mixing of intralayer excitons due to the tunneling and symmetry-breaking perturbations.

Let us first include exciton tunneling as a whole described by the matrix element $T_{\rm B}$ in the Hamiltonian, which tends to form symmetric and antisymmetric combinations of B-excitons  as well as symmetry breaking between the MLs in bilayer resulting in the splitting of B-excitons with $E_{\rm B,1} - E_{\rm B,2} = \delta E_{\rm B}$:
\begin{equation}
\label{Ham:ext:B}
\mathcal H = \begin{pmatrix}
E_{\rm IE} - \mu F_z & T &0 &0\\
T^* & E_{\rm B,1} & 0 & T_{\rm B}\\
0 & 0 & E_{\rm IE} + \mu F_z & T\\
0 & T_{\rm B}^* & T^* & E_{\rm B,2}
\end{pmatrix}.
\end{equation}
We focus at the reasonable relation between the parameters of the system with the electric field being low as compared with the broadening of the indirect exciton, $|\mu F_z| \ll \Gamma_{\rm IE} \ll \Delta$, but, on the other hand, sufficiently high so that $|\mu F_z| \gg |TT_{\rm B}/\Delta|$, $\delta E_{\rm B} |T|^2/\Delta^2$. Under these approximations the $T_{\rm B}$ and $\delta E_{\rm B}$ do not affect the IE exciton and its susceptibility is given by Eq.~\eqref{chi:IE}. 

The effective Hamiltonian describing the doublet of B-excitons in the second-order in $T$ and lowest order in $\mu F_z$, $\delta E_B$ reads
\begin{equation}
\label{Ham:ext:BB}
\mathcal H = \begin{pmatrix}
E_{\rm B,1}+ \mu F_z \frac{|T|^2}{\Delta^2} + \frac{|T|^2}{\Delta}&  T_{\rm B}\\
T_{\rm B}^* & E_{\rm B,2} - \mu F_z \frac{|T|^2}{\Delta^2} + \frac{|T|^2}{\Delta}
\end{pmatrix}.
\end{equation}
Furthermore, in a weak asymmetry limit where $|T_B| \gg \delta E_B, |\mu F_z|{|T|^2}/{\Delta^2}$, the Hamiltonian~\eqref{Ham:ext:BB}  in the basis of $(|\mathrm B_1\rangle\pm|\mathrm B_2\rangle)/\sqrt{2}$ states  transforms to (we assume that the diagonal terms $|T|^2/\Delta$ are included in $E_{\rm B}$)
\begin{equation}
\label{Ham:ext:BB:1}
\mathcal H = \begin{pmatrix}
E_{\rm B}+ T_{\rm B}&  \delta E_{B}+\mu F_z \frac{|T|^2}{\Delta^2} \\
\delta E_B +  \mu F_z \frac{|T|^2}{\Delta^2}  & E_{\rm B} - T_{\rm B}
\end{pmatrix},
\end{equation}
whose eigenstates are:
\begin{align}
|\mathrm B_+\rangle: \quad E_{\mathrm B_+} = E_{\rm B} + T_{\rm B}, \\ 
 |\mathrm B_+\rangle  = \frac{|\mathrm B_1\rangle + |\mathrm B_2\rangle}{\sqrt{2}} + \frac{\delta}{2T_{\rm B}}\frac{|\mathrm B_1\rangle - |\mathrm B_2\rangle}{\sqrt{2}}, \nonumber\\
|\mathrm B_-\rangle: \quad E_{\mathrm B_-} = E_{\rm B} - T_{\rm B}, \\
 |\mathrm B_+\rangle  = \frac{|\mathrm B_1\rangle - |\mathrm B_2\rangle}{\sqrt{2}} - \frac{\delta}{2T_{\rm B}}\frac{|\mathrm B_1\rangle + |\mathrm B_2\rangle}{\sqrt{2}}, \nonumber
\end{align}
with $\delta = \delta E_{\rm B} + \mu F_z |T|^2/\Delta^2$. As a result, the non-linear susceptibility in the spectral range of B-excitons reads
\begin{multline}
\label{susc:B:ext}
\chi = \frac{\delta}{2T_B} CM_1M_2 \times \\
\left(\frac{1}{2\hbar\omega - E_{\rm B} - T_{\rm B} + \mathrm i \Gamma_B} - \frac{1}{2\hbar\omega - E_{\rm B} + T_{\rm B} + \mathrm i \Gamma_B}\right)
\\
\approx \left(\delta E_{\rm B} + \frac{\mu F_z |T|^2}{\Delta^2} \right) \frac{{CM_1M_2}}{(2\hbar\omega - E_{\rm B} +  \mathrm i \Gamma_{\rm B})^2-T_{\rm B}^2}
\end{multline}

Very similar treatment can be made for A-exciton assuming that it is mixed with the IE(B) exciton, which follows from the spin conservation rule. Making natural changes: $T_{\rm B}$ to the tunneling matrix element of the A-exciton $T_{\rm A}$, $\Delta$ to the splitting $\Delta'=E_\text{A}-E_\text{IE(B)}$,  and $\delta E_{\rm B}$ to the A-exciton splitting $E_{\rm A,1} - E_{\rm A,2} = \delta E_{\rm A}$, we obtain
\begin{equation}
\label{susc:A:ext}
\chi_\text{A} = \delta' {C'M_1M_2\over (2\hbar\omega - E_{\rm A} + \mathrm i \Gamma_{\rm A})^2{-}T_{\rm A} ^2}.
\end{equation}
Here $\delta'=\delta E_{\rm A} + \mu F_z |T_{\rm A}|^2/{\Delta'}^2$. It follows from Eq.~\eqref{susc:A:ext} that the effect of the electric field on A-exciton is weaker than on the IE. It follows from atomistic calculations that the A-excitons can be mixed with interlayer exciton IE$_1$ via the spin-flip electron tunneling. This effect can be included in our model in a same way, resulting in susceptibilities similar to Eqs.~\eqref{chi:IE} and \eqref{chi:B} derived for the mixing of the IE- and B-excitons.\\

\textbf{Acknowledgements.---} Toulouse acknowledges funding from ANR 2D-vdW-Spin, ANR MagicValley, ANR IXTASE, ANR HiLight, ITN 4PHOTON Marie Sklodowska Curie Grant Agreement No. 721394 and the Institut Universitaire de France. 
 Growth of hexagonal boron nitride crystals was supported by the
Elemental Strategy Initiative conducted by the MEXT, Japan ,Grant Number
JPMXP0112101001,  JSPS KAKENHI Grant Number JP20H00354 and the
CREST(JPMJCR15F3), JST.
 M.M.G. and L.E.G. acknowledge the RFBR and CNRS joint project 20-52-16303. L.E.G. was supported by the Foundation for the Advancement of Theoretical Physics and Mathematics ``BASIS''.\\


\end{document}